\documentclass[twocolumn]{aastex631}

\usepackage{amsmath}

\received{May 27, 2021}
\revised{Feb 17, 2022}
\accepted{Mar 19, 2022}   
\submitjournal{ApJ}

\shorttitle{Coherent differntial imaging on speckle area nulling}
\shortauthors{Nishikawa}

\graphicspath{{./}{figures/}}

\begin{document}

\title{The coherent differential imaging on speckle area nulling (CDI-SAN) method for high-contrast imaging under speckle variation }

\correspondingauthor{Jun Nishikawa}
\email{jun.nishikawa@nao.ac.jp}

\author[0000-0001-9326-8134]{Jun Nishikawa}
\affiliation{TMT project, National Astronomical Observatory of Japan,
2-21-1 Osawa, Mitaka, Tokyo, 181-8588, Japan}
\affiliation{Department of Astronomical Science, School of Physical Sciences, The Graduate University for Advanced Studies (SOKENDAI), 2-21-1 Osawa, Mitaka, Tokyo, 181-8588, Japan}
\affiliation{Astrobiology Center, 2-21-1 Osawa, Mitaka, Tokyo, 181-8588, Japan}
\nocollaboration{1}

\begin{abstract}
Differential imaging is a postprocessing method to obtain high contrast, often used for exoplanet searches. The coherent differential imaging on speckle area nulling (CDI-SAN) method was developed to detect a faint exoplanet lying beneath residual speckles of a host star. It utilizes image acquisitions faster than the stellar speckle variation synchronized with five shapes of a deformable mirror repeatedly. By using the only the integrated values of each of the five images and square differences for a long interval of observations, the light of the exoplanet could be separated from the stellar light. The achievable contrast would reach to almost the photon-noise limit of the residual speckle intensities under appropriate conditions. The CDI-SAN can be applied to both ground-based and space telescopes. 
\end{abstract}

\keywords{Exoplanet detection methods(489) --- Direct imaging(387) --- Coronagraphic imaging(313) --- Exoplanets(498)} 

\section{INTRODUCTION} \label{sec:intro}
Various high-contrast imaging methods have been developed to detect the light of exoplanets for characterization. Planet intensities observed from reflected light have been estimated to be about $10^{-8}$ to $\sim 10^{-10}$ of the intensity of its host star. To achieve such high-contrast, diffracted starlight in the telescope should be removed by a coronagraph, and residual starlight should be suppressed by wave-front control. Observed images would be often reduced by postprocessing techniques called differential imaging, which would use two or more images at different wavelengths, polarizations, or orientation angles to further suppress the residual starlight speckles. 

Real-time focal-plane wave-front sensing and control methods utilized speckle intensities with some coherent-light modulations and nulled the speckles to produce high-contrast regions mainly in laboratory experiments leading to space telescopes 
\citep[e.g.,][]{2004ApJ...615..562G, 2004SPIE.5487.1330T, 2006ApJ...638..488B, 2006dies.conf..553B, 2011SPIE.8151E..10G, 2015OptRv..22..736O}. In this process, speckle intensities were derived separately from the incoherent component (planet light). Coherent differential imaging (CDI) is a method that uses the same principle in postprocessing \citep[e.g.,][]{2018SPIE10703E..1UJ}. 
In a ground-based telescope using wave-front control by a wave-front sensor, the coherent-light technique was applied to reduce quasi-static diffracted light and speckles \citep[e.g.,][]{2014PASP..126..565M,2017MNRAS.464.2937B}.  

Recently, a fast operation in the CDI method mentioning the potential of long exposure using a self-coherent camera was proposed \citep{2018AJ....156..106G}. In the present paper, a fast synchronized detection process and long-exposure technique based on the electric field conjugation \citep[EFC;][]{2011SPIE.8151E..10G} and the speckle area nulling \citep[SAN;][]{2015OptRv..22..736O} algorithms is presented; this is a kind of CDI, and we call it the CDI on speckle area nulling (CDI-SAN) method. The principle behind CDI-SAN will be described in section 2. Numerical simulations and discussions will be shown in section 3. Conclusions will be described in section 4.  

\section{Principle behind CDI-SAN} \label{sec:prin}
The EF at the focal plane can be modulated by controlling a deformable mirror (DM) at the pupil plane. Let the intensity without modulation at a detector pixel in the final focal plane be
\begin{equation}
\label{eqI0}
I_0=I_s+I_p, 
\end{equation}
then those with some modulations can be written as
\begin{equation}
\label{eqI12}
\left\{
\begin{array}{l} 
I_1^+=\left|E_s+\Delta E_1\right|^2+I_p \\
I_1^-=\left|E_s-\Delta E_1\right|^2+I_p \\
I_2^+=\left|E_s+\Delta E_2\right|^2+I_p \\
I_2^-=\left|E_s-\Delta E_2\right|^2+I_p \\ 
\end{array}
\right. 
\end{equation}
\begin{figure*}[htbp]
 \gridline{	\fig{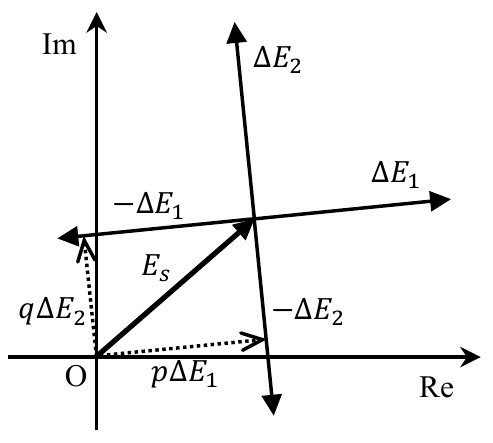}{0.28\textwidth}{(a)}   \fig{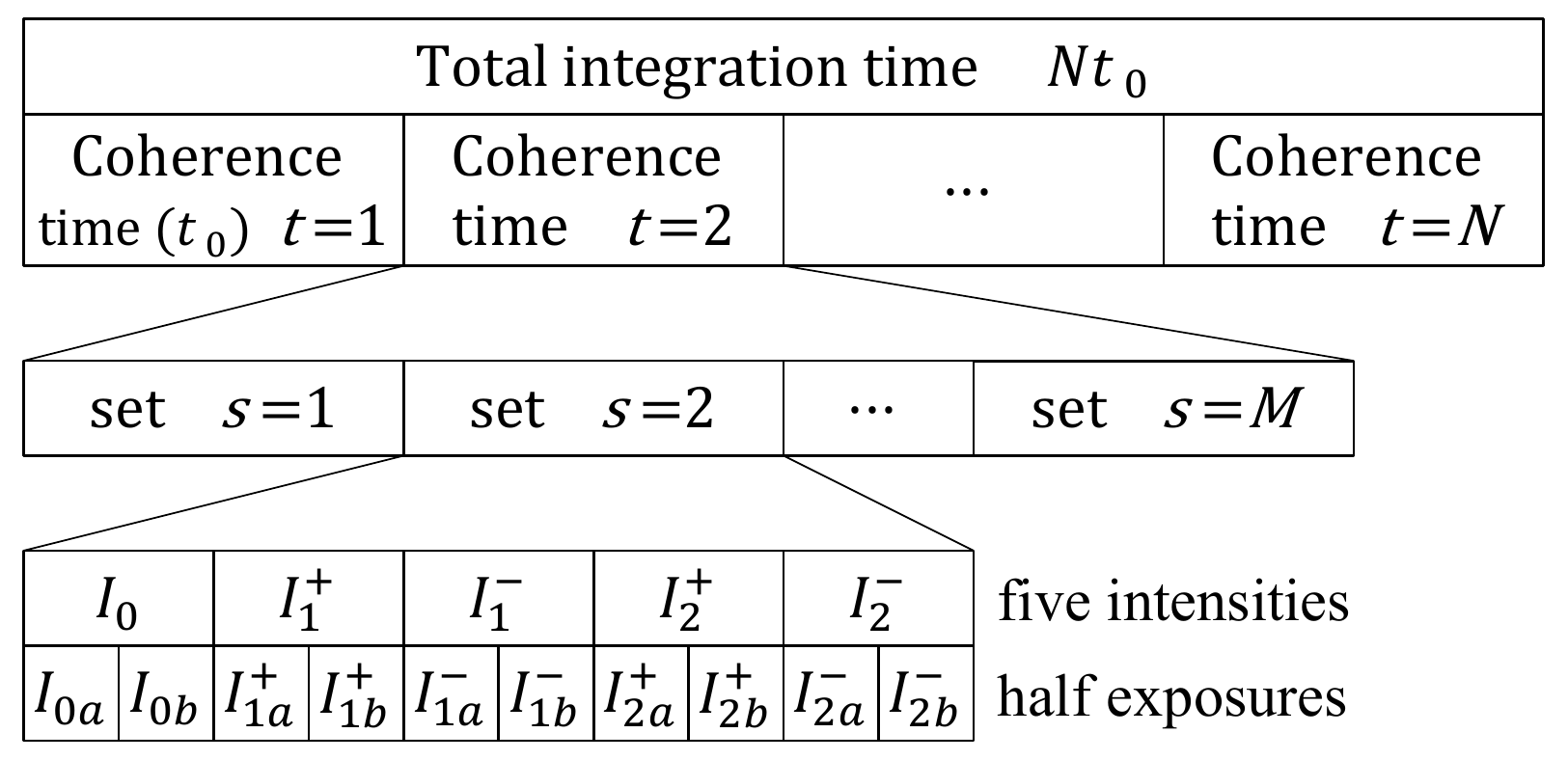}{0.5\textwidth}{(b)} }
 \caption{Measurement configurations. (a) Electric field at a pixel and (b) data acquisition sequence.}
 \label{EFdata}
\end{figure*}
where $I_s$ and $E_s$ are the intensity and an EF of the speckle with zero modulation, respectively, i.e., $I_s=\left|E_s\right|^2$, $I_p$ is the planet intensity including other light incoherent to the starlight, and $\Delta E_1$ and $\Delta E_2$ are the nonzero modulated EFs. The pair-wise modulations used in the EFC and the SAN, $\pm \Delta E_1$ and $\pm \Delta E_2$, were generated by the $\pm \rm{sin}$ and $\pm \rm{cos}$ shapes of the wave-front at the pupil plane, respectively, to make $\Delta E_1$ and $\Delta E_2$ perpendicular in the complex plane (Figure~\ref{EFdata}~(a)). Let us define a set as a sequence of the five intensities $(I_0, I_1^+, I_1^-, I_2^+$ and $I_2^-$; see Figure~\ref{EFdata}~(b)). All intensities are treated as contrasts, normalized by the peak of a point spread function (PSF) of a telescope, in the present paper. After solving the simultaneous equation of the Equations (\ref{eqI0}) and (\ref{eqI12}) on $E_s$, the DM is controlled to delete $E_s$ by producing $-E_s$ at many pixels to make a dark area---thus called dark-field control---and a final measurement is made without the modulations after reducing $I_s$ to the lowest possible level. The details of the control are not the interest of the present paper, although the control is recommended (not required) in parallel with the CDI-SAN observation. The solution of $E_s$ was not resolved into real and imaginary parts as in the EFC but into $\Delta E_1$ and $\Delta E_2$, which were used in the SAN and the present method (Figure~\ref{EFdata}~(a)) as 
\begin{equation}
\label{eqEs}
\left\{
\begin{array}{l} 
E_s =p\Delta E_1+q\Delta E_2 \\
p =\frac{\Delta E_1\bullet E_s}{\left|\Delta E_1\right|^2\ }=\frac{\left(I_1^+-I_1^-\right)}{2\left(I_1^++I_1^--2I_0\right)} \\
q =\frac{\Delta E_2\bullet E_s}{\left|\Delta E_2\right|^2\ }=\frac{\left(I_2^+-I_2^-\right)}{2\left(I_2^++I_2^--2I_0\right)},\\
\end{array}
\right. 
\end{equation}
 where $\bullet $ is the inner product when the EF is considered as a two-dimensional vector in Cartesian coordinates. 

The five intensity measurements should be made within a timeframe where the variation of $E_s$ (hereafter $dE_s$) is negligibly small, or the simultaneous equations will be disturbed. In the case of a ground-based telescope, atmospheric turbulence uncompensated by an adaptive optics {(AO)} system would be the cause of $dE_s$ at the tens of millisecond scale. In a future space telescope, deformation of the telescope optical structure might produce $dE_s$ under a long exposure of minutes and hours aiming at detections of faint exoplanets. 

To overcome this problem, the CDI-SAN repeatedly utilizes measurements of the set that are faster than $dE_s$ and integrates the data for a long interval well beyond the time constant of $dE_s$ in postprocessing  (Figure~\ref{EFdata}~(b)). Let us calculate the average of zero-modulation intensities as 
\begin{equation}
\label{eqI0A}
\left< I_0 \right> = \left< I_s \right> +I_p 
\end{equation}
and those with some modulations as
\begin{equation}
\label{eqI12A}
\left\{
\begin{array}{l} 
\langle I_1^+ \rangle =\langle \left|E_s+\Delta E_1\right|^2 \rangle +I_p \\
\langle I_1^- \rangle =\langle \left|E_s-\Delta E_1\right|^2 \rangle +I_p \\
\langle I_2^+ \rangle =\langle \left|E_s+\Delta E_2\right|^2 \rangle +I_p \\
\langle I_2^- \rangle =\langle \left|E_s-\Delta E_2\right|^2 \rangle +I_p , \\ 
\end{array}
\right. 
\end{equation}
where $\langle \rangle$ means the average of all observed sets. Assuming that $\Delta E_1$, $\Delta E_2$ and $I_p$ are constant, the relations 
\begin{equation}
\label{eqDEsq}
\left\{
\begin{array}{l} 
\left|\Delta E_1\right|^2\ ={\left(\langle I_1^+\rangle +\langle I_1^-\rangle  -2\langle I_0\rangle \right)/2} \\
\left|\Delta E_2\right|^2\ ={\left(\langle I_2^+\rangle +\langle I_2^-\rangle  -2\langle I_0\rangle \right)/2} \\
\langle\left|\Delta E_1\bullet E_s\right|^2\rangle\ =\langle\left(I_1^+\ -I_1^-\right)^2\rangle/16 \\
\langle\left|\Delta E_2\bullet E_s\right|^2\rangle\ =\langle\left(I_2^+\ -I_2^-\right)^2\rangle/16  \\
\end{array}
\right. 
\end{equation}
were derived. Then, we obtain the averaged speckle intensity as
\begin{eqnarray}
\label{eqIsA}
\langle I_{s} \rangle & = & \frac{\langle\left|\Delta E_1\bullet E_s\right|^2\rangle}{\left|\Delta E_1\right|^2\ } 
+\frac{\langle\left|\Delta E_2\bullet E_s\right|^2\rangle}{\left|\Delta E_2\right|^2\ } \nonumber \\
 & = & \frac{\langle\left(I_1^+-I_1^-\right)^2\rangle }{8\left( \langle I_1^+ \rangle + \langle I_1^- \rangle -2 \langle I_0 \rangle \right) } \nonumber \\ 
& + & \frac{\langle\left(I_2^+-I_2^-\right)^2\rangle }{8\left( \langle I_2^+ \rangle + \langle I_2^- \rangle -2 \langle I_0 \rangle \right) }, 
\end{eqnarray}
where the square of the absolute value of $E_s$ in Equation~(\ref{eqEs}) and $\Delta E_1\bullet \Delta E_2=0$ were used. Finally, from Equation~(\ref{eqI0A}), we estimate the unknown planet intensity by removing the averaged speckle intensity $\langle I_s \rangle$ from the averaged zero-modulation intensity $\langle I_0 \rangle$ as 
\begin{eqnarray}
\label{eqIp1}
I_{p1} =  \langle I_0 \rangle - \frac{\langle\left(I_1^+-I_1^-\right)^2\rangle }{8\left( \langle I_1^+ \rangle + \langle I_1^- \rangle -2 \langle I_0 \rangle \right) } \nonumber \\
- \frac{\langle\left(I_2^+-I_2^-\right)^2\rangle }{8\left( \langle I_2^+ \rangle + \langle I_2^- \rangle -2 \langle I_0 \rangle \right) } 
\end{eqnarray}
which is the essence of CDI-SAN, an exact solution of the planet intensity just derived using only the average values over a long interval, i.e., the five averaged intensities and the two averaged square differences. 

{Using Equations~(\ref{eqIsA}) and (\ref{eqIp1}) need not be limited to the star and planet light in varying wave-front environments. This refers to methods to estimate $\langle I_s \rangle$ (averaged coherent-light intensity) and $I_p$ (incoherent intensity) from the integrated values when additional coherent EFs of $\Delta E_1$ and $\Delta E_2$ are available. }
Equations~(\ref{eqIsA}) and (\ref{eqIp1}) are the exact solutions and can provide a contrast of infinity, i.e., $I_{p1}$=$I_p$ even for $10^{-100}$, if $E_s$ does not change within a set (perfect correlation), where different $E_s$ can be acceptable for different sets. 
{Rather less correlation---randomness---with changing $E_s$ is the problem. So, in the worst case, the performance of the equations themselves under the no-correlation condition (random change of $E_s$) will be good to study first.} 
Intensity changes owing to fast $dE_s$ during a set produce errors in the simultaneous equations, which can be considered to act as a kind of noise (hereafter speckle variation noise, SVN). 

Other than that, the photon shot noise (hereafter PSN) and readout noise (hereafter RON or $R$) of a detector will affect to the intensity measurements significantly, when photon flux becomes low. Then, the Equation~(\ref{eqIp1}) could be developed into another essential equation for CDI-SAN with bias corrections for the mean square difference values in the numerators as 
\begin{eqnarray}
\label{eqIp2}
I_{p2} =  \langle I_0 \rangle 
- \frac{\langle\left(I_1^+-I_1^-\right)^2\rangle -V_1^+\ -V_1^-}{8\left( \langle I_1^+ \rangle + \langle I_1^- \rangle -2 \langle I_0 \rangle \right) } \nonumber \\
- \frac{\langle\left(I_2^+-I_2^-\right)^2\rangle -V_2^+\ -V_2^-}{8\left( \langle I_2^+ \rangle + \langle I_2^- \rangle -2 \langle I_0 \rangle \right) } ,
\end{eqnarray}
where $V_1^+, V_1^-, V_2^+$, and $V_2^-$ are the noise variances of $I_1^+, I_1^-, I_2^+$, and $I_2^-$, respectively. The noise variances would be estimated by additional mean square differences as
\begin{equation}
  V_X^Y= \langle\left(I_{Xa}^Y-I_{Xb}^Y\right)^2\rangle/4 \ \ \ \ \left(_X=1\ \rm or\ 2,\ \it _Y=+\ \rm or\ -\right)   
\end{equation}
where $I_{Xa}^Y$ and $I_{Xb}^Y$ are two half-{exposures} of $I_X^Y$ (see Figure~\ref{EFdata}~(b)), which satisfy 
\begin{equation}
 I_X^Y=\left(I_{Xa}^Y+I_{Xb}^Y\right)/2 \ \ \ \  \left(_X=1\ \rm or\ 2,\ \it _Y=+\ \rm or\ -\right),  
\end{equation}
 where the denominators, 4 and 2, are required because the half-{exposures} are the contrast normalized by the PSF's peak. It is expected that the averaging process for $I_{p1}$ or $I_{p2}$ for a long interval will produce a high-contrast image by suppressing these noises. The amplitude of $\Delta E_1$ and $\Delta E_2$ should be adjusted carefully because the solutions diverge if $\langle I_1^+ \rangle + \langle I_1^- \rangle -2 \langle I_0 \rangle$ or $\langle I_2^+ \rangle + \langle I_2^- \rangle -2 \langle I_0 \rangle $ becomes close to zero deu to noise. Because 
{the noise behaviors} 
in the Equations (\ref{eqIp1}) and (\ref{eqIp2}) are not simple, we need numerical simulations to know their characteristics. 

{After revealing how the equations themselves performe, we need to investigate some issues in implementation. We know how to generate $\Delta E_1$ and $\Delta E_2$ for a wide area in the focal plane using a DM operation with the shape formed by the sum of many sin and cosine waves with various wavenumbers \citep[][]{2011SPIE.8151E..10G, 2015OptRv..22..736O}; however, the target area should be limited to suppress the maximum DM stroke when $\Delta E_1$ and $\Delta E_2$ become large amplitudes for bright speckle conditions. Here, the orthogonality of $\Delta E_1$ and $\Delta E_2$ should also be considered. The contrast estimation with an appropriate spatial distribution of $\langle I_s \rangle$ is interesting. $E_s$ correlations after $t_0$ as in the atmospheric wave-front variations, are welcome and would be advantageous relative to the no-correlation conditions. } 

\section{Simulations and discussions}
In this section the CDI-SAN method will be examined using numerical simulations. 
{The equations are solvable for each pixel so that the performance of the solutions of the equations themselves are independent of the spatial distributions of $\langle I_s \rangle$, $\Delta E_1$, and $\Delta E_2$. Therefore, it is effective to estimate the standard deviation (SD) and rms (root-mean-square) from the measured pixel intensities calculated in parallel at many pixels within a defined area assuming uniform $\langle I_s \rangle$, $\Delta E_1$, and $\Delta E_2$ to know the noise level of a pixel after an observation. It would not be good to estimate them by combining them with many nonuniformities, which would make the performance of the equations themselves hard to understand.}

A measurement sequence is shown in Figure~\ref{EFdata}~(b). The set consisted of the five intensities, and they are divided into two half-{exposures} suffixed with $a$ and $b$; the set then included 10 half-{exposures}. Let us introduce a time constant, $t_0$, and have $E_s$ change to a new value every $t_0$. The number of sets obtained during $t_0$ is $M$, and the observation was made up to $N \times t_0$. 
{As mentioned in Section 2, a fixed $E_s$ does not generate any errors in the equations and can be removed exactly using the present method; then, the strong-correlation condition would be an advantage and the SVN would become small. But in the simulation here, $E_s$ after $t_0$ was assumed to be 100\% independent, which would be the worst condition for the equations. } 

The $E_s$ was generated as a focal-plane EF by a fast Fourier transform {(FFT)} of a pupil function with a wave-front error, the diameter of which was 128 in a $256\times256$ array, where the EF without the wave-front error (the PSF of the pupil) was subtracted as a coronagraph.  
{The pupil function was unity within the diameter of the circular aperture and zero outside. The wave-front phase error (hereafter $\phi$, omitting the pupil-plane coordinates, generated in the full of array) was assumed to have a spectrum of a flat-power up to a radius of 80 followed by a power law of -4 outside, and random arguments, where the complex conjugate numbers were forced at axisymmetric positions to make the phase error a real number. Its absolute amplitude was then adjusted to have an SD of $10\sqrt{C}$ waves in the pupil, where $C$ was the desired mean contrast of a rectangular extracted region (explained later) and $C=1\times10^{-5}$ (hereafter case~1, for a ground-based telescope). The phase error was applied to the pupil function as $e^{i\phi}$. A wave-front amplitude error was not considered. The subtraction of the PSF after the FFT was equivalent to a subtraction of unity in the pupil before the FFT. 
Under a small wave-front-error approximation, the residuals after the unity subtraction would become $i\phi$ in the pupil area, and then $I_s$ would have the same intensity distribution as the spectrum of $\phi$, flat within the radius of 80 pixels and not depending on the distance from the center. Therefore, we can extract anywhere from the flat-spectrum area to study the present method, except for the correlated pixels at the axisymmetric positions. } 
A focal-plane area of {$100\times50$} pixels 
{$(-50\le x\le49, 11\le y\le60)$} was extracted. 
{At this point, the average of $I_s$ within the rectangular area was about $C$, then it was adjusted to $C$ by multiplying a correction ratio to $E_s$. In addition to case~1, $C=1\times10^{-9}$ for a space telescope will be discussed as case~2. 
}

\begin{figure*}
\gridline{\fig{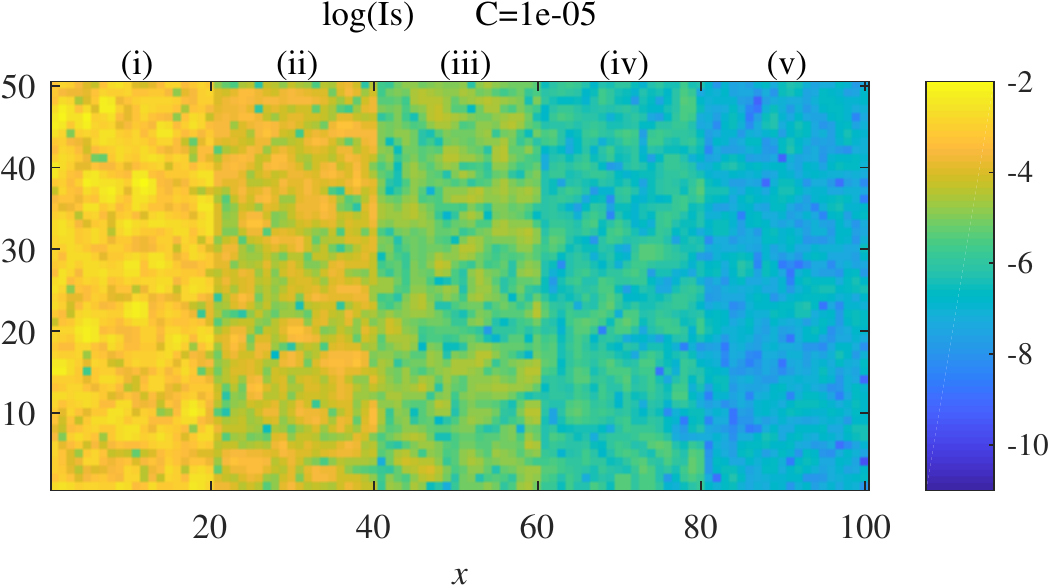}{0.45\textwidth}{(a)}   \fig{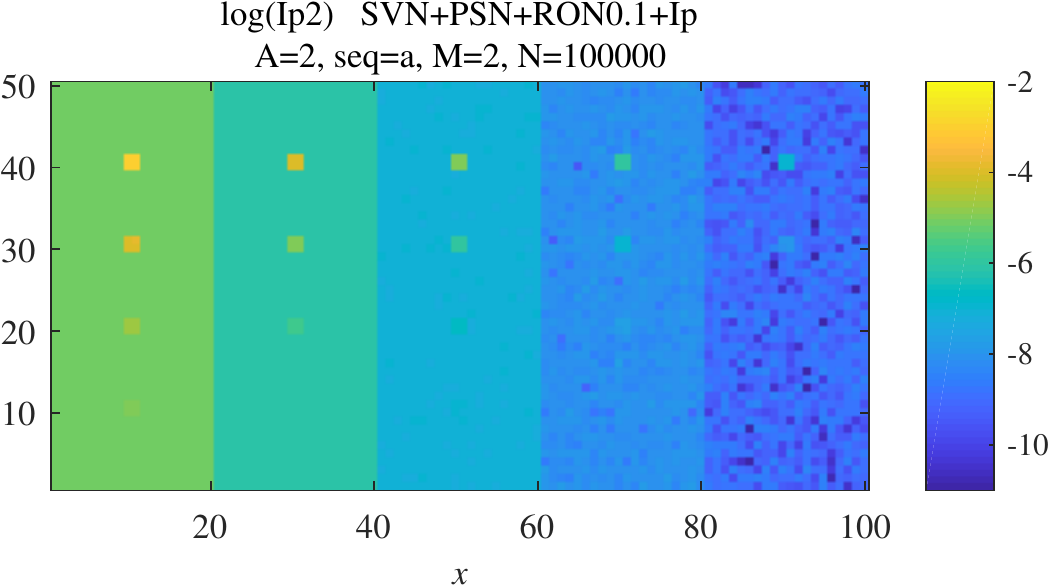}{0.45\textwidth}{(b)} }
\gridline{\fig{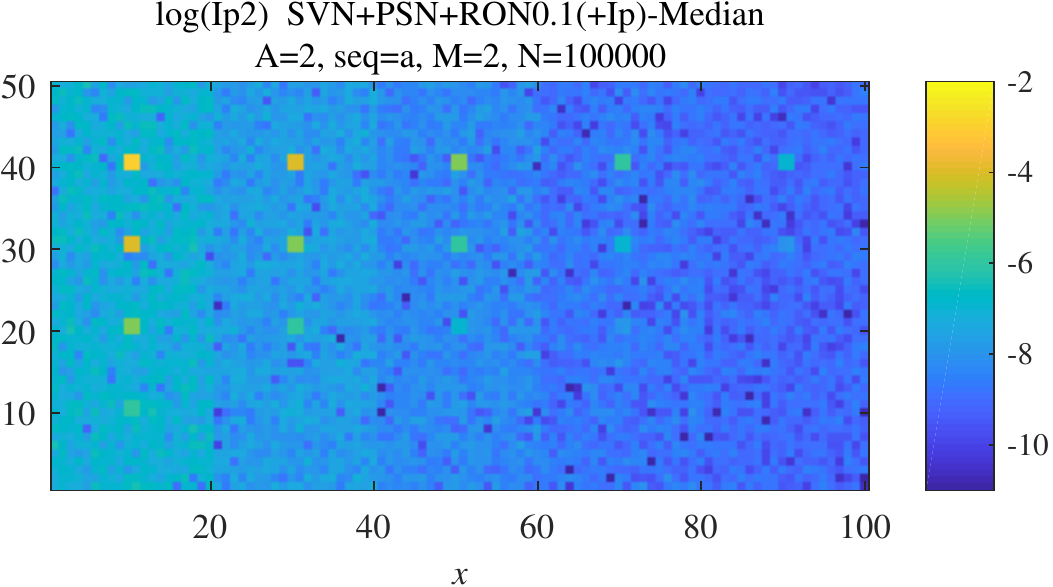}{0.45\textwidth}{(c)}   \fig{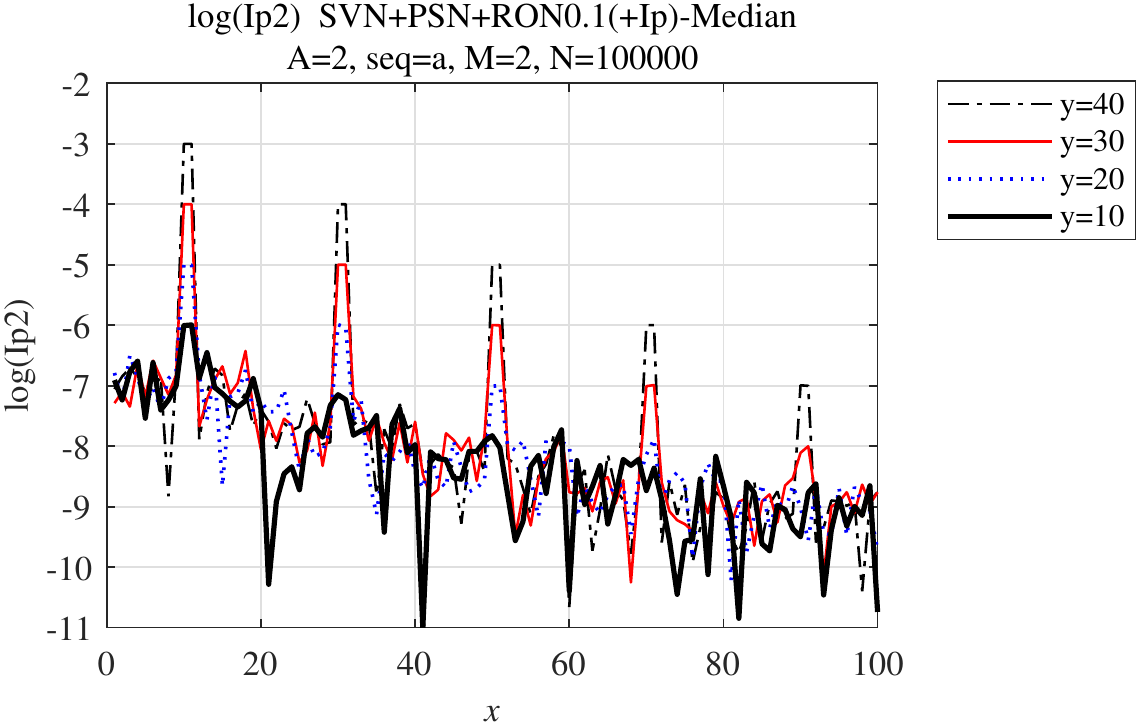}{0.45\textwidth}{(d)} }
 \caption{Simulation images and profiles. (a) An example of the initial focal-plane speckle image with $C=1\times{10}^{-5}$. (b) $I_{p2}$ with $M$=2 and $N$=100,000 estimated with SVN, PSN, and RON=0.1. Artificial planets whose intensities were $BC (\sim \langle\langle I_0 \rangle\rangle), BC/10, BC/100$, and $BC/1000$ at $y=40-41, 30-31, 20-21$, and $10-11$, respectively, were included at two central pixels on $x$ in each region. (c) $I_{p2}$ image bias-subtracted from (b). (d) Horizontal profiles of (c) at $y=40$(dot-dashed black line), 30 (solid red line), 20 (blue dot line), and 10 (thick black line). The values in all panels are shown in log absolute. }
\label{Ip2image}
\end{figure*}

A new independent $E_s$ was generated at $t_0$ intervals and nonlinearly interpolated for the $10M$ half-{exposures} between the intervals using weights of $(1-w)/W$ and $w/W$ for the previous wave-front error and the new one, respectively, where $w=k/10M$ ($k=1$ to $10M$) and $W=(w^2+\left(1-w\right)^2)^{1/2}$. 
The area was, as shown in Figure~\ref{Ip2image}~(a), divided into five regions every 20 pixels in the $x$ direction, (i), (ii), (iii), (iv), and (v), considering a regional contrast coefficient (hereafter $B$) of 100, 10, 1, 0.1, and 0.01, respectively, to cover wide range contrasts at a time. Now $\sqrt{B}$ was multiplied to $E_s$, and the contrast was 
{forced to} stepwise $BC$. Figure~\ref{Ip2image}~(a) is an example $I_s$ of one exposure without modulation and noise. 

A photon flux (hereafter $P$) per $t_0$ at $C$ must be introduced, and $P$=100 was adopted, then the flux $BP$=10,000, 1000, 100, 10, and 1 at each region. The photon flux depends on a target star's magnitude, a telescope diameter, and observation conditions, and $P$=100 would be roughly obtained, e.g., at \mbox{GJ 411} (M2V, $m_J$=4.2) using a 30m telescope, a wavelength of 1.2$\mu$m, and $t_0$=36 ms in case~1 or at a solar-like star of 10 pc distance using a 9 m telescope, a wavelength of 500 nm, and $t_0$=3600 s in case~2, under the common conditions of a total photon detection efficiency of 0.2, an acceptance of a quarter energy of the PSF at a pixel, and a band width of 2\%. 

For the modulation EF, constant values of $\Delta E_1=A\sqrt{I_n} $ and $\Delta E_2=\Delta E_1i$ were adopted for each region, where $A$ was a modulation amplitude coefficient, $i$ was an imaginary unit, and $I_n$ was an expected noise intensity estimated by  $I_n=(B^2P^2+BP+10MR^2)^{1/2}C/P$ considering the SVN, the PSN, and the RON. Here $A$=2 was adopted as it showed the best contrast among $A= 1, 2, 3, 4, 5,$ and 10 in most conditions. 
{The contrast becomes worse with $A$ less than 1, which will be discussed later. }

Half-{exposures} were made including the PSN and the RON, and the data were integrated. An image of $I_{p2}$ for the case~1 was shown in Figure~\ref{Ip2image}~(b) under the conditions of $M$=2, $R$=0.1, and $N$=100,000, which corresponded to a 3600 s observation when $t_0$=36 ms. We could recognize bias intensities in bright regions, and then the image appeared to be improved by subtracting the bias intensities calculated using the median values at individual regions as shown in Figure~\ref{Ip2image}~(c) and at four horizontal profiles plotted in Figure~\ref{Ip2image}~(d). The artificial planets observed in the figures were included in the half-{exposures}. The brightest planet in each region had an intensity of $BC$ and the best appeared above the noise levels were down to $BC/1000$, $BC/100$, and $BC/10$, at (i) and (ii), (iii), and (iv) and (v), respectively. 

\begin{figure*}
\gridline{ \fig{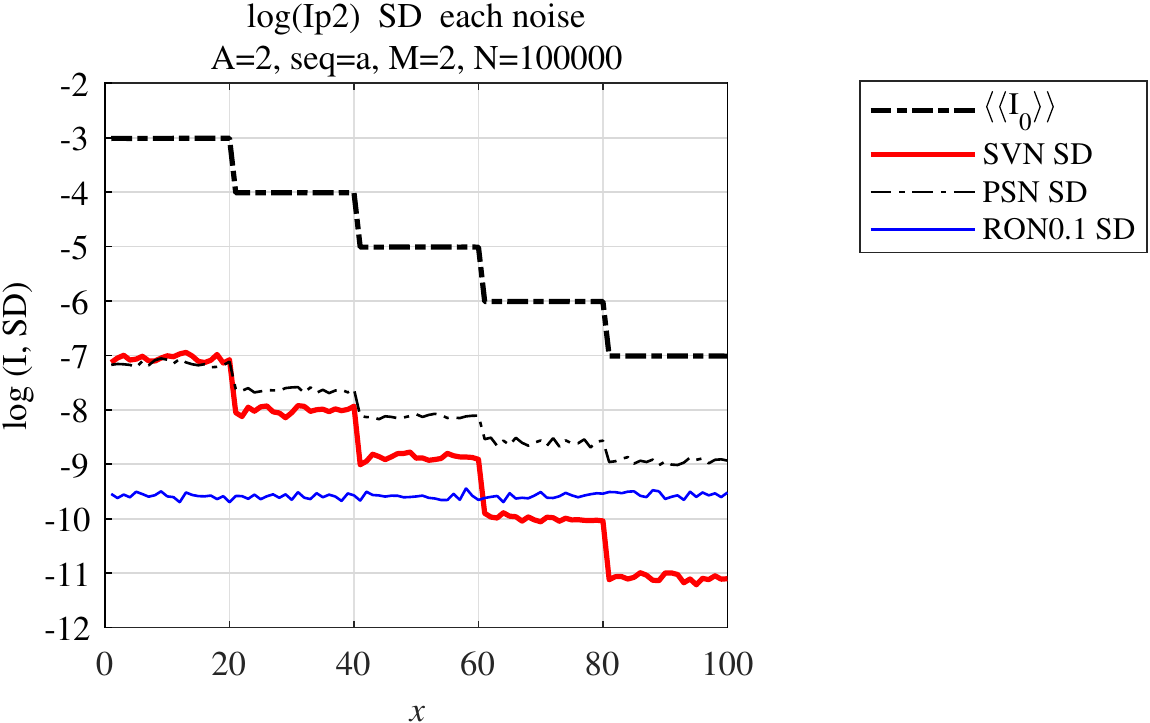}{0.45\textwidth}{(a)}  \fig{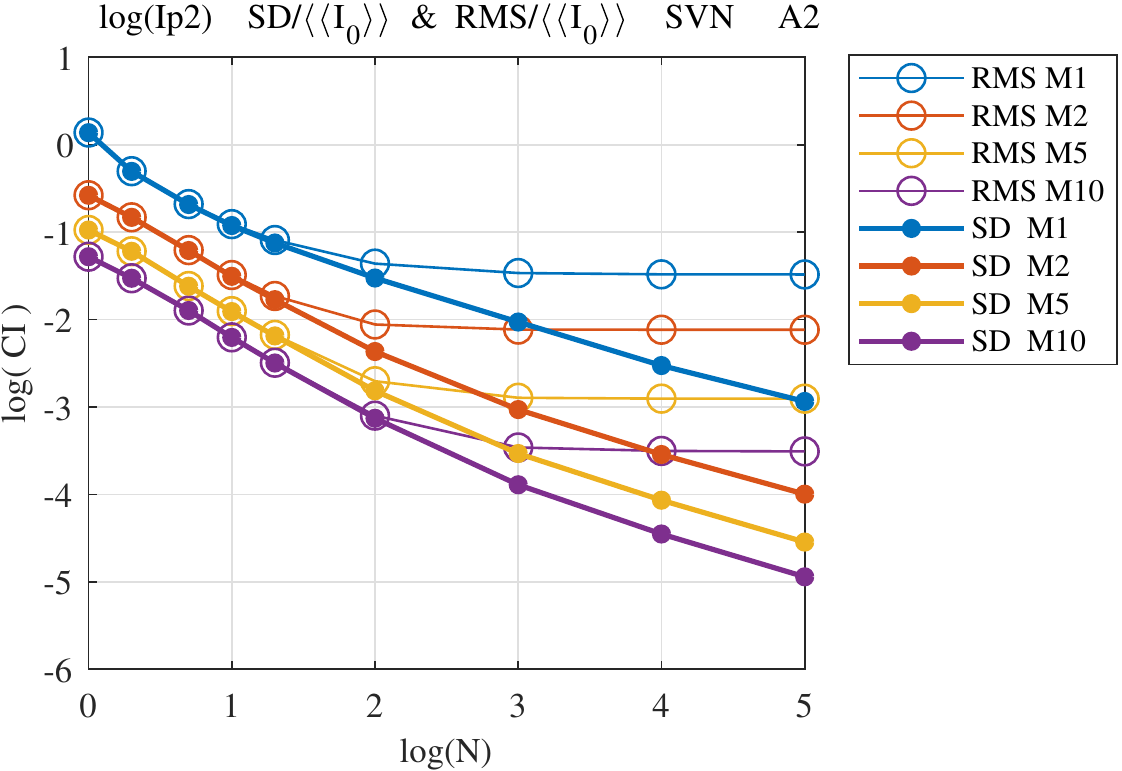}{0.45\textwidth}{(b)}  }
\gridline{ \fig{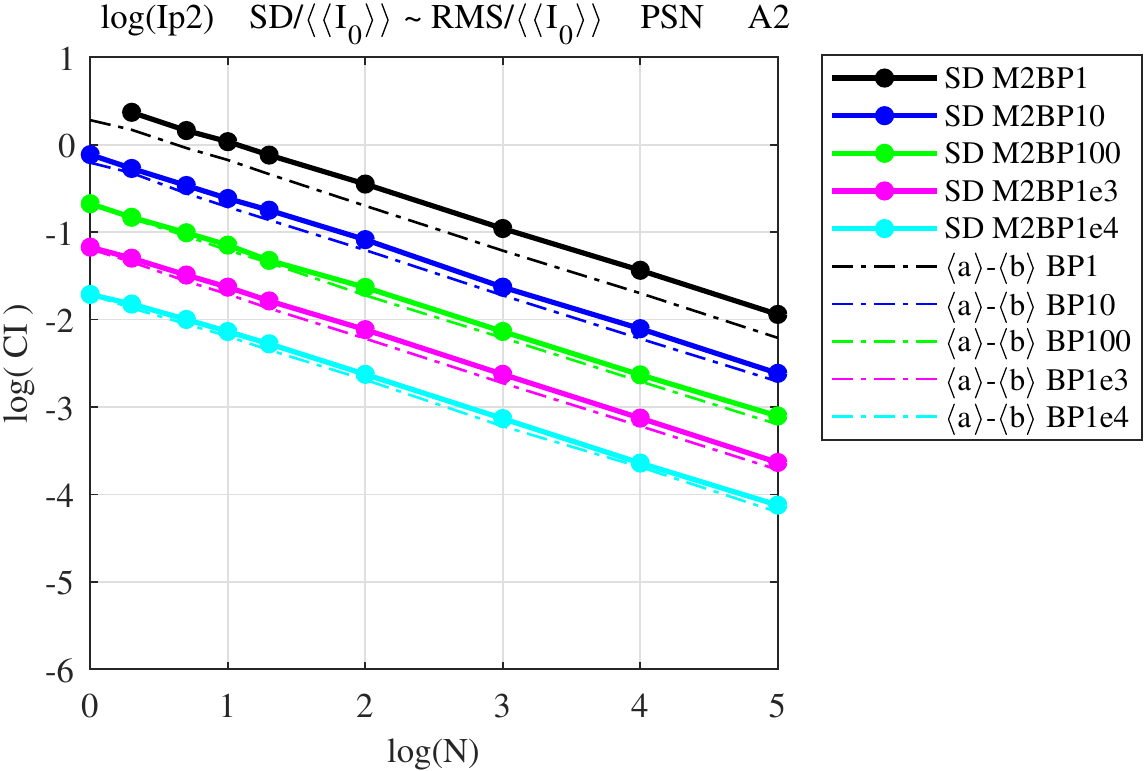}{0.45\textwidth}{(c)}  \fig{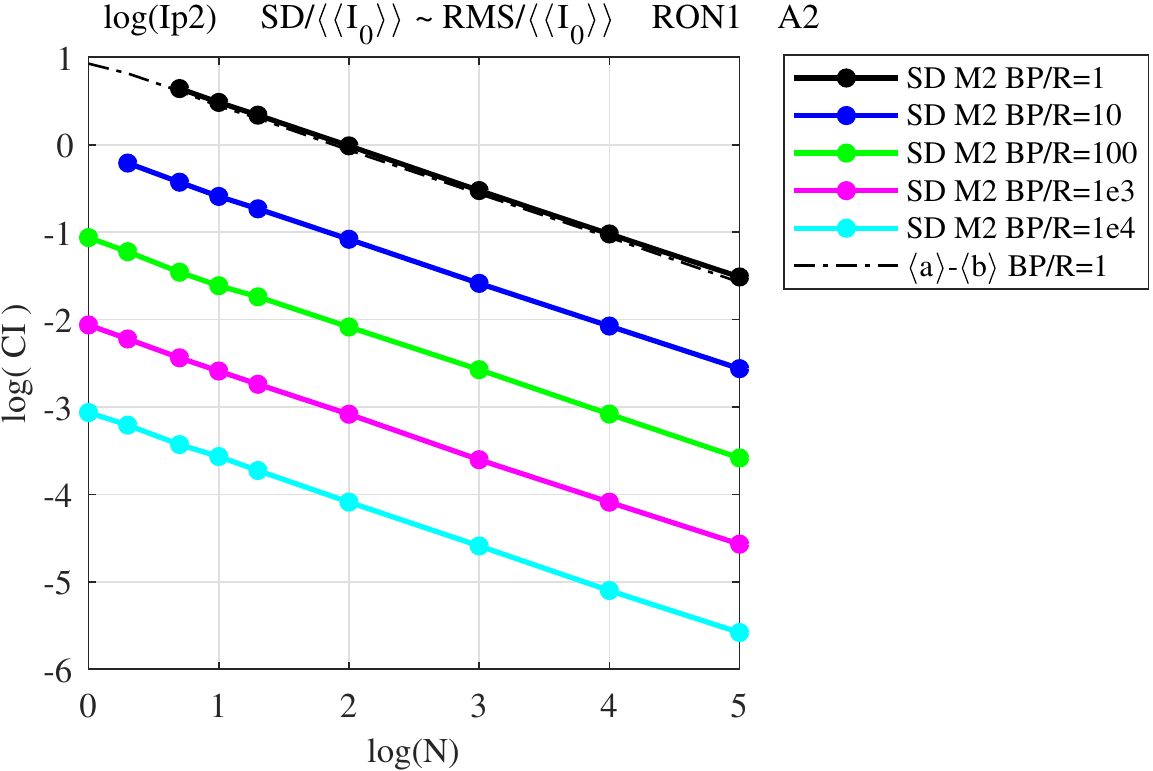}{0.45\textwidth}{(d)}  }
\caption{Effects of the individual three noises in $I_{p2}$. The ordinate values were shown in log scale. (a) SD$_{x}$ with $M$=2 and $N=$100,000. Tthick red line: SVN, dotted-dashed black line: PSN, solid blue line: RON(=0.1), thick dotted-dashed black line: $\langle\langle I_0 \rangle\rangle$, original contrast as a reference. (b) SVN. Thick lines with filled circles: SD/$\langle\langle I_0 \rangle\rangle$, and solid lines with open circles: rms/$\langle\langle I_0 \rangle\rangle$, for $M$=1, 2, 5, and 10 from top to bottom in both lines. (c) PSN with $M$=2. Thick lines with filled circles: SD/$\langle\langle I_0 \rangle\rangle$, and dotted-dashed lines: PSN-limit lines calculated from $(\langle a\rangle-\langle b\rangle)/\langle\langle I_0 \rangle\rangle$. (d) RON with $M$=2 and $R$=1. Thick lines with filled circles: SD/$\langle\langle I_0 \rangle\rangle$, and  dotted-dashed line: RON-limit line with $BP/R$=1 using $(\langle a\rangle-\langle b\rangle)/\langle\langle I_0 \rangle\rangle$. The lines in (c) and (d) were for $BP$=1, 10, 100, 1000, and 10,000 from top to bottom. 
}
\label{Ip2CI}
\end{figure*}

\begin{figure*}
\gridline{	\fig{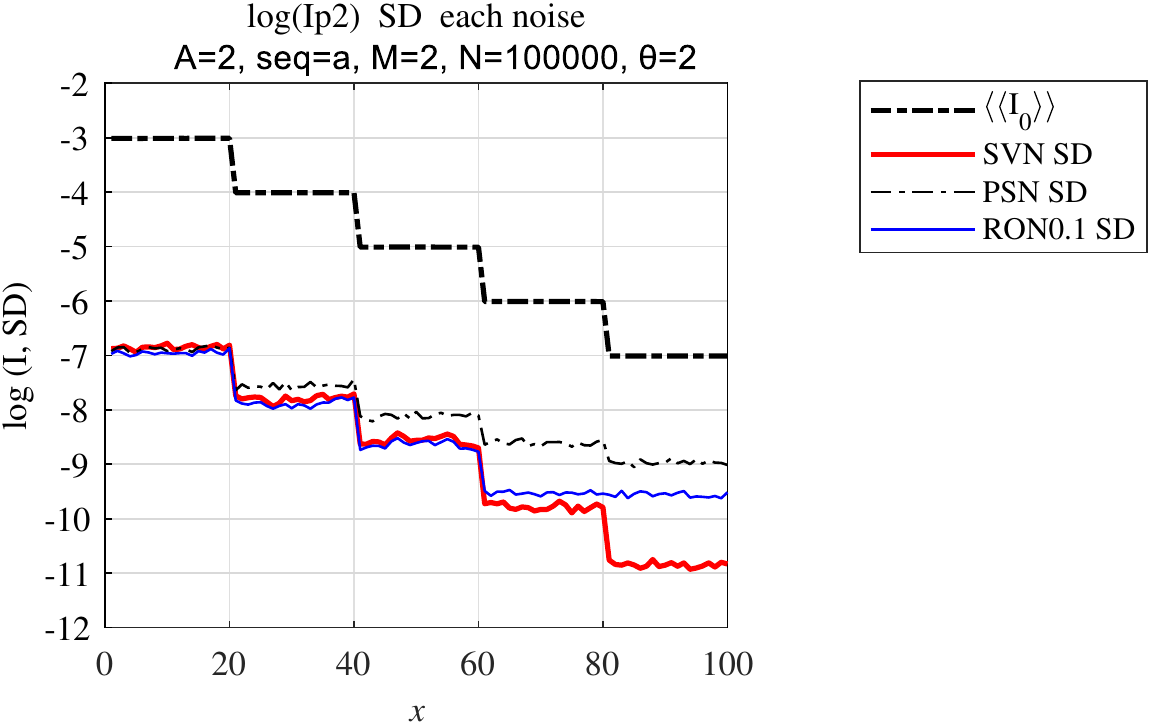}  {0.43\textwidth}{(a)}    \fig{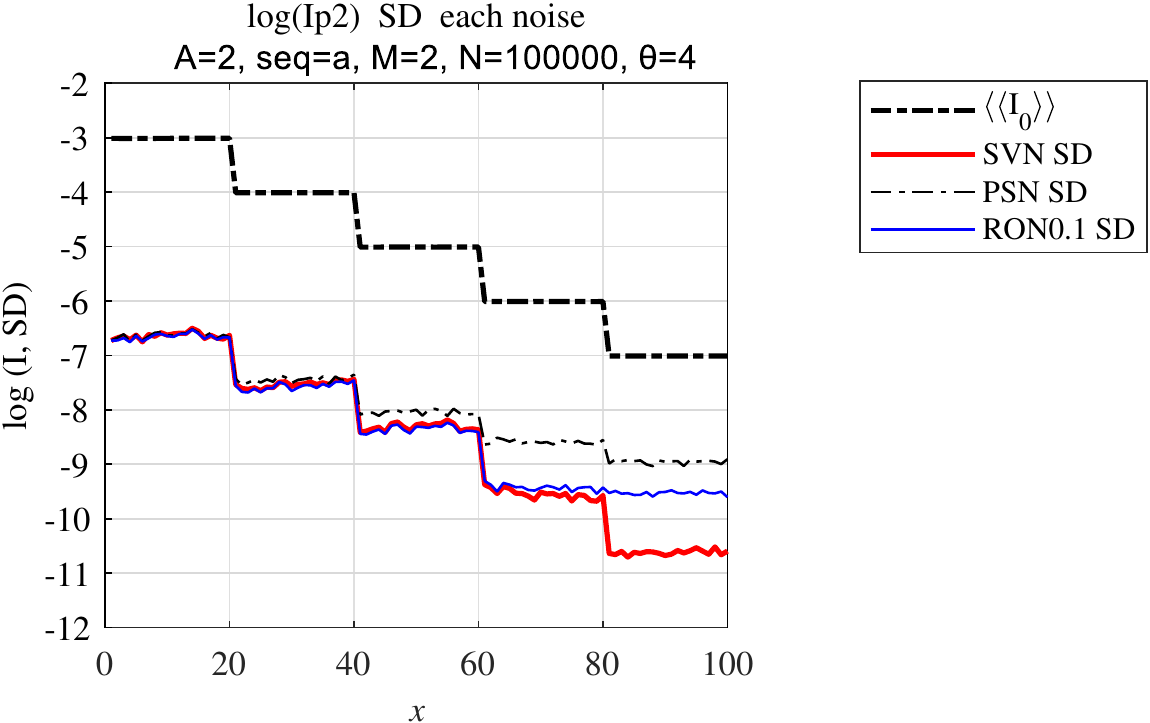}  {0.43\textwidth}{(b)}  }
\caption{The SD of $I{p2}$ for each noise as in Figure \ref{Ip2CI}(a) calculated with the relative argument error $\theta$. (a) $\theta=2^\circ$, and (b) $\theta=4^\circ$. 
}
\label{Thetadep}
\end{figure*}

Next, the $I_{p2}$ was recalculated by adding either the SVN, the PSN, or the RON individually without the planets to separate their contributions to the noise level, and an SD was estimated. An SD of 50 pixels in $y$ at each $x$ (hereafter SD$_{x}$) is shown in Figure~\ref{Ip2CI}~(a), where ${\langle\langle I_0 \rangle\rangle}$ indicates the original contrast in reference to the average of ${\langle I_0 \rangle}$ for the corresponding pixels. The effect of the RON was successfully lower than that of the PSN in all regions. It was found that the effect of the SVN was well suppressed to the same level as the PSN at (i) and lower in other regions, which meant that the present method with $M$=2 could overcome the d$E_s$ problem under the conditions evaluated here. The same figures for various $N$ can be seen in Appendix~\ref{VariousN}. 

An SD at each region was then estimated and normalized by ${\langle\langle I_0 \rangle\rangle}$, expressed as SD/${\langle\langle I_0 \rangle\rangle}$ hereafter, which indicated a contrast improvement (hereafter CI), where ${\langle\langle I_0 \rangle\rangle}$ is an average of ${\langle I_0 \rangle}$ at the corresponding region. 
An rms/${\langle\langle I_0 \rangle\rangle}$ was also estimated as another CI to obtain the bias intensity, where rms was from each region. 
The SD of $\langle I_0 \rangle$ is equal to $\langle\langle I_0 \rangle\rangle$ and $BC$ statistically because the speckle intensity followed a chi-square distribution with two degrees of freedom. 

Figure~\ref{Ip2CI}~(b) showed the SD/${\langle\langle I_0 \rangle\rangle}$ with only the SVN, where the values are the average for all regions this time because they did not depend on the region. 
The larger $M$ and the larger $N$ have better contrast at least within the ranges shown here. The slope at $N\geq1000$ was almost $1/\sqrt{N}$, and better slopes were obtained at $N\leq100$ for $M\geq2$. The achievable contrast with $M\geq2$ is obviously better than that with $M$=1 at $N\geq100$. $M\geq2$ is recommended if other conditions permit. The rms/${\langle\langle I_0 \rangle\rangle}$ is also plotted in Figure~\ref{Ip2CI}~(a), where its saturations are found at $N\geq100$ indicating bias intensities produced by the SVN. That was why bias intensities were observed in bright regions in Figure~\ref{Ip2image}~(b). The bias intensity was positive and decreased with a large $M$. If the bias intensity can be removed by a smooth surface at the focal plane as seen in Figure~\ref{Ip2image}~(c) and (d), we do not have to use a large $M$ to suppress the bias. A small $M$ is effective to slow the data acquisition speed and to reduce parts of the SVN and the RON as discussed later. Therefore, $M$=2 was used in the evaluations hereafter. 

Figure~\ref{Ip2CI}~(c) showed SD/${\langle\langle I_0 \rangle\rangle}$ with only the PSN for the five regions with a different photon flux of $BP$ under $M$=2. Here the $dE_s$ during a set stayed zero, i.e., the wave front changed only at the beginning of each set, to get no SVN even under the wave-front change condition. As reference, the SDs of ${(\langle a \rangle-\langle b \rangle)/\langle\langle I_0 \rangle\rangle}$ are also shown, where $\langle a \rangle$ and $\langle b \rangle$ are averages of all the half-{exposures} suffixed by $a$ and $b$, respectively. They lie at $2/\sqrt{BPN}$, which are considered PSN limits of ideal differential images with halves. The SD/${\langle\langle I_0 \rangle\rangle}$ of $I_{p2}$ showed a good performance with a small increase of only 1.2 of the PSN limit at $BP\geq100$ between $10\le N\le100,000$, where almost no {\it M} dependence was observed. At low-light levels, the increase is still only about 1.3 and 1.8 times at $BP$=10 and $BP$=1, respectively, and gradually rises with $M$. The rms is almost the same as SD which meant that the bias was well removed by the correction terms in Equation~(\ref{eqIp2}). 

Figure~\ref{Ip2CI}~(d) shows the SD/${\langle\langle I_0 \rangle\rangle}$ with only the RON for the five regions under $M$=2, where $R$=1 was used to see whether it is worse than $R$=0.1. The lines at $5\le N\le100,000$ were almost coincident with the SD of ${(\langle a \rangle-\langle b \rangle)/\langle\langle I_0 \rangle\rangle}$ calculated as in the PSN, which lies on ideal RON limits from the function $2R\sqrt{10M}/BP\sqrt{N}$. Even with $R$=0.1 and 10 and $M$=1-5, the results were close to the function (0.8-1.2 times), e.g., the condition $BP/R$=1/0.1 was comparable to 10/1. 
The rms was almost the same as the SD, and again, the bias was well removed by the correction terms in the Equation~(\ref{eqIp2}). The successful bias corrections for the RON and the PSN were the reason why Figure~\ref{Ip2image}~(b) showed no bias in the low-light regions. 
The CIs of $I_{p1}$ without the bias corrections, Equation~(\ref{eqIp1}), are mentioned in Appendix~\ref{Ip1} because they were not the point of the present paper. 

Now the CIs of the CDI-SAN method against the three noises, which would be considered independent, have been resolved. Also, the CIs in the high-contrast case~2 ($P$=100 at $C=1\times10^{-9})$ turned out to be the same as those in case~1 described above (Figure~\ref{Ip2CI}~(b), (c), and (d)) because they depend on $P$ instead of on $C$, although $N$ should just be $\la 100$ (see the SD$_x$ in Figure~\ref{Ip2CIN10} (a) and (b) of Appendix~\ref{VariousN}). An important result was that CDI-SAN could approach the natural limit of the PSN by suppressing the SVN and the RON with appropriate $M$ and $R$ under the available conditions of $(B)P$ and $N$. 

\begin{figure}
\gridline{	\fig{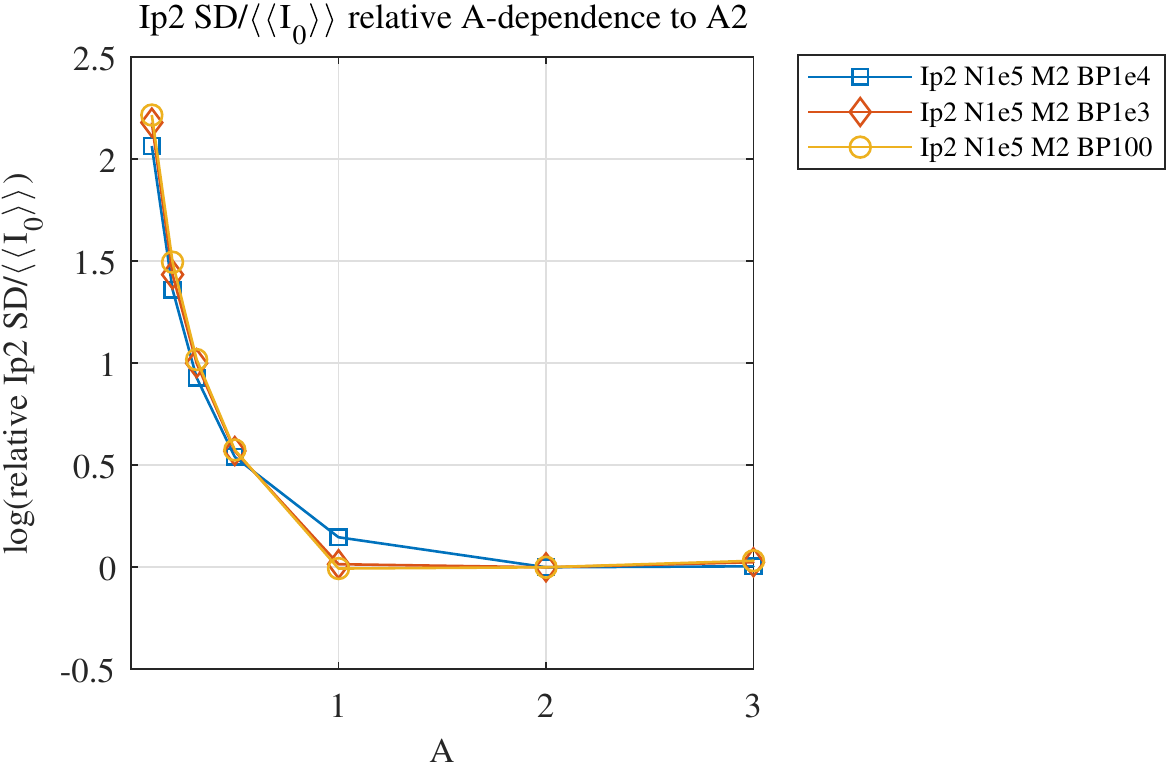}  {0.46\textwidth}{}  }
\caption{$A$ dependenc of the SD/${\langle\langle I_0 \rangle\rangle}$ of $I_{p2}$ normalized by the value at $A$=2, for $M$=2 and $N=$100,000. Ssquare: $BP$=10,000, diamond: $BP$=1000, and circle: $BP$=100.
}
\label{Adep}
\end{figure}

\begin{figure*}
\gridline{	\fig{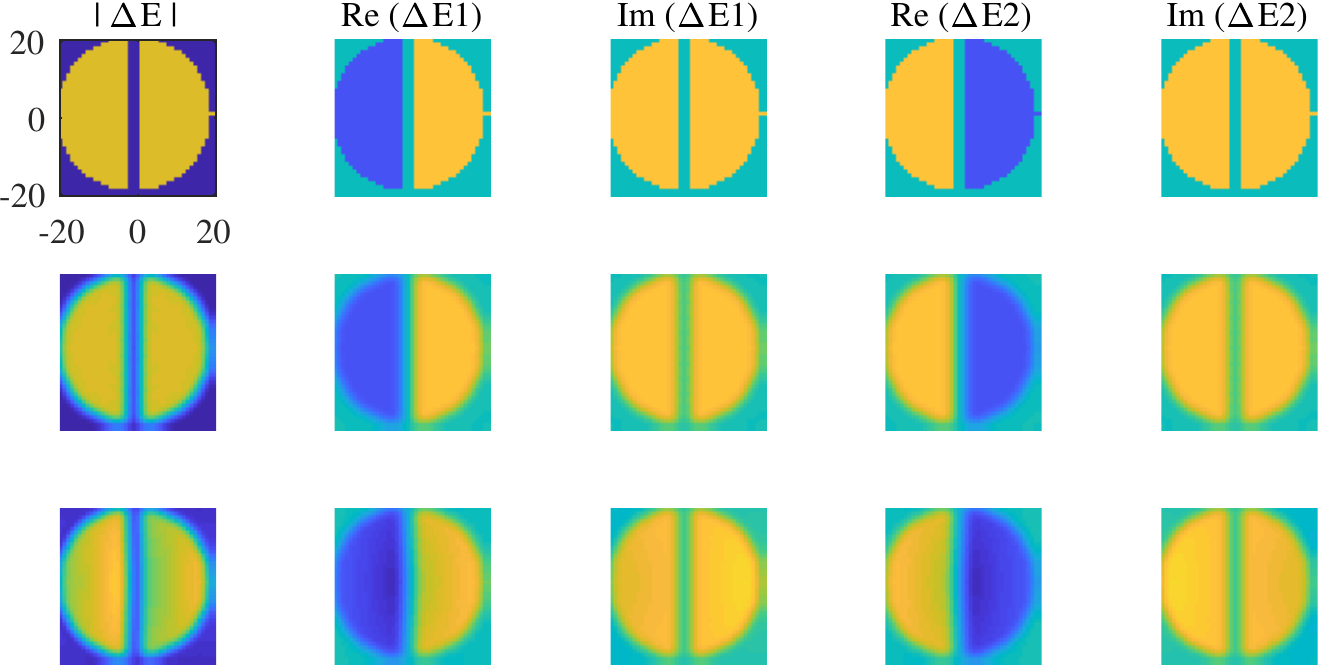}  {0.7\textwidth}{}  } 
\caption{Generation of the modulation EF in arbitrary units. The upper panels are the initial desired EFs, the middle panels are the second desired EFs after the pupil function and the apodization are operated on the pupil plane, and the bottom panels are the EFs generated by the operation $e^{i\phi}$ in the pupil and FFT. }
\label{ModEF}
\end{figure*}

\begin{figure}
\gridline{	\fig{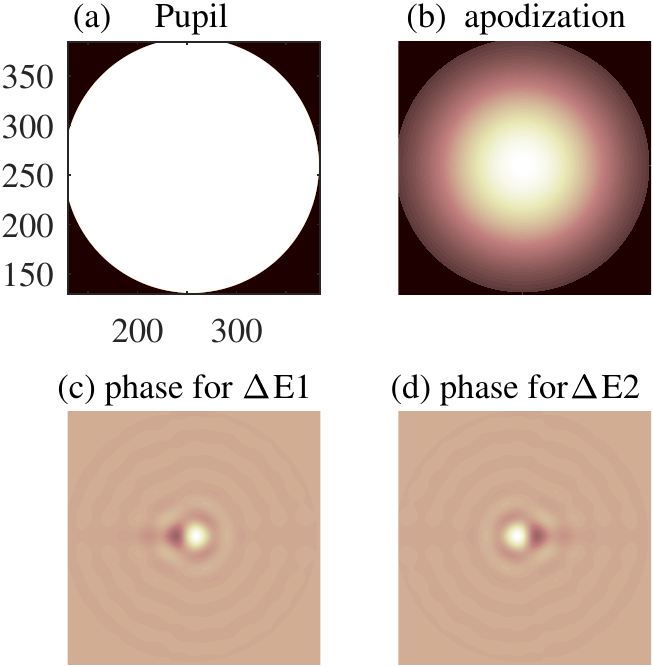}  {0.32\textwidth}{}  } 
\caption{(a) Pupil function, (b) apodization function, (c) wave-front to generate $\Delta E_1$, and (d) wave-front for $\Delta E_2$. }
\label{PupilApodi}
\end{figure}

{
The orthogonality of the $\Delta E_1$ and $\Delta E_2$ is important to achieve high contrast. 
An error of the relative argument between them, hereafter $\theta$, would produce an increase in the SD. Assuming that arg($\Delta E_1)=0^\circ$ and arg($\Delta E_2$)=arg($\Delta E_1$)+$90^\circ+\theta$, the SD/${\langle\langle I_0 \rangle\rangle}$ of $I_{p2}$ was calculated and shown in Figure~\ref{Thetadep}. It was found that the $\theta$ increased the SDs for all the noises. The log(CI) was limited to about -3.6 as seen at the region (i) for $\theta=4^\circ$. The orthogonality of the $\Delta E_1$ and $\Delta E_2$ should be considered according to the raw contrast and the target contrast.   A similar result was confirmed with arg($\Delta E_1$)=$45^\circ$ and arg($\Delta E_2$)=$135^\circ$ which will be used later. 
 
When $\theta$ is known given by a model, it is possible to use a solution without the orthogonality assumption---a free-argument solution---which is shown in Appendix~\ref{FreeArgument}. It would be useful if the difference between the model and real optics is small. 
}

{
The modulation amplitude coefficient, $A$, would affect the SD. 
The $A$-dependent SD/${\langle\langle I_0 \rangle\rangle}$ of $I_{p2}$ relative to $A$=2 is shown in Figure~\ref{Adep} for $M=$2 and $N=$100,000.  It strongly affects the SD when it is less than 1, which depends on about $1/A^2$, while almost no effect is confirmed between 1 and 10. We recommend using $A=1$ to $\sim2$ because a very bright modulation would cause a problem in real optics. \\ 
The above describes the characteristics of the solution of the equation itself. 
}

{Next, generating the modulation EFs with the DM in real optics will be discussed. A candidate of a modulation wave front (hereafter MWF), i.e., phase map, at a pupil plane can be calculated by the inverse FFT of the desired EF, $\Delta E_1$ or $\Delta E_2$, at a target area in the focal plane \citep[e.g.,][]{2011SPIE.8151E..10G}. 
Let us determine the amplitude and argument of the initial desired modulation EF, and then calculate the real and imaginary parts  (see Figure \ref{ModEF} top panels). 
The EF is the negative conjugate at each axisymmetric position. 
Let us adopt arg($\Delta E_1$)=$45^\circ$ and arg($\Delta E_2$)=$135^\circ$ to retain good orthogonality even at a large amplitude of the modulation EF. 
After an inverse FFT operation, only the imaginary part is generated. A pupil function and an apodization function are operated on (see Figure \ref{PupilApodi}, top panels), where the latter of which is used to get a flatter amplitude of the modulation EF, and the candidate value of the MWF is obtained in the imaginary part (see Figure \ref{PupilApodi}, bottom panels). 
The FFT operation of this function generates the second desired modulation EF (see Figure \ref{ModEF}, middle panels). Finally, the candidate value is used as the phase of the MWF, and FFT is used to obtain the (final) modulation EF (see Figure \ref{ModEF}, bottom panels). The focal-plane modulation EF can be generated by an FFT of the pupil function with the MWF and subtracting the unaberrated pupil function. When the MWF is very small, the (final) modulation EF is almost equal to the second desired modulation EF. 
}

\begin{figure*}
\gridline{	\fig{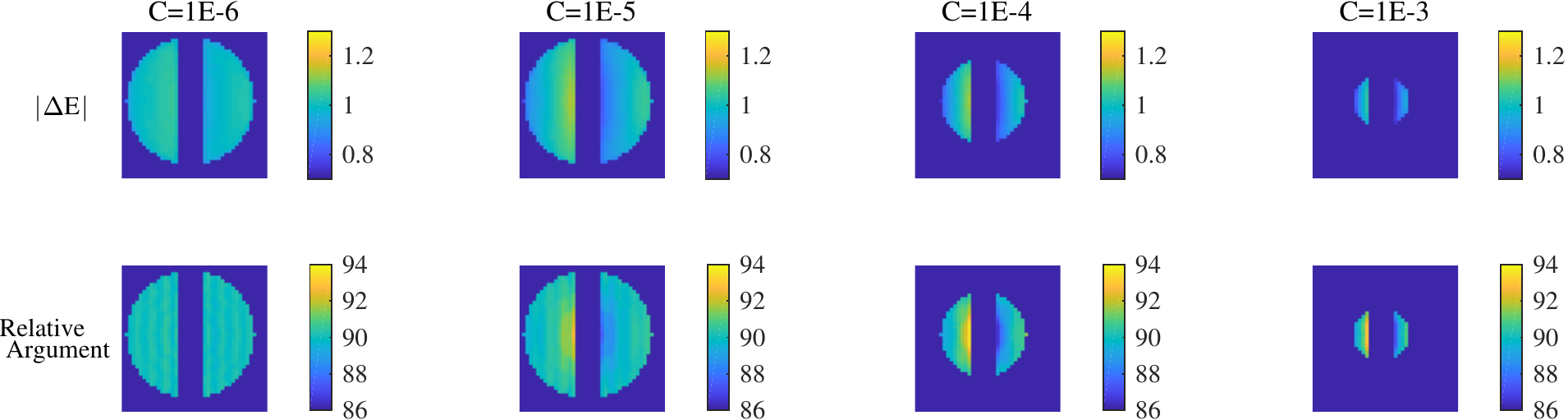}  {0.9\textwidth}{}  } 
\caption{Modulation EFs generated by modulation wave-fronts. The top panels are the amplitude and the bottom panels are the relative argument between $\Delta E_1$ and $\Delta E_2$.  Desired contrasts are $10^{-6}, 10^{-5}, 10^{-4}$, and $10^{-3}$ from left to right. The modulation area has a radius of 20 pixels (10 $\lambda/D$) and $\left| x \right| \ge 2$ pixels (1 $\lambda/D$) for the two darker contrast cases. The color map shows the evaluation area, which is 1 $\lambda$/D smaller than the initial desired modulation area. Detailed values are listed in Table 1. } 
\label{ModEF2}
\end{figure*}

\begin{deluxetable*}{cccccc}
\tablenum{1}
\tablecaption{Modulation EF parameters shown in Figure \ref{ModEF2} }
\tablewidth{0pt}
\tablehead{
\colhead{Contrast} & \colhead{Outer Radius} & \colhead{Inner $x$} & \colhead{$\left|E\right|$} & \colhead{Relative}   & \colhead{Wave-Front Phase}   \\
\colhead{} & \colhead{Evaluation (Initial)} & \colhead{Evaluation (Initial)} & \colhead{Relative to Initial} & \colhead{Argument} &{}  \\
\colhead{} & \colhead{($\lambda/D$)} & \colhead{($\lambda$/D)} & \colhead{min-max} & \colhead{min-max ($^\circ$)}& \colhead{min-max (radian)}   \\
}
\startdata
$10^{-6}$ & 9 (10) & 2 (1) & 0.86--1.04 & 89.5--90.3 & -0.11--0.20\\
$10^{-5}$ & 9 (10) & 2 (1) & 0.77--1.14 & 87.1--90.6 & -0.37--0.71 \\
$10^{-4}$ & 6 (7) & 2 (1) & 0.74--1.11 & 86.6--91.3 & -0.52--0.95 \\
$10^{-3}$ & 3.9 (4.9) & 2 (1) & 0.74--1.03 & 87.1--90.8 & -0.76--1.30 \\
\enddata
\end{deluxetable*}

{
The generated modulation EFs whose desired intensities are $10^{-6}, 10^{-5}, 10^{-4}$, and $10^{-3}$ are shown in Figure \ref{ModEF2} and characteristic values are listed in Table 1. These are the candidates with relative arguments within $90 \pm 3^\circ.5$, taking into account the discussion above, and whose amplitude degradation was better than 0.74. Here an FFT array of $512\times512$ was provided and the diameter was 256 pixels; then, 2 pixels/$\lambda/D$ was obtained. The modulation area had a radius of 20 pixels (10 $\lambda/D$) and $\left| x \right| \ge 2$ pixels (1 $\lambda/D$) for the two darker contrast cases, and others can be seen in Table 1. The modulation EF was evaluated 2 pixels inside the modulated area in all cases, although pixels outside the evaluation area were not absolutely unusable. Even for the large, nearly 1 radian wave-front phase, which is a difficult condition to approximate in the imaginary part, the relative argument   
between $\Delta E_1$ and $\Delta E_2$ stayed around 90 $^\circ$ by setting arg($\Delta E_1$)=$45^\circ$ and arg($\Delta E_2$)=$135^\circ$, although they changed by more than a dozen degrees at maximum from the initial desired modulation EF. Setting arg($\Delta E_1$)=$0^\circ$ and arg($\Delta E_2$)=$90^\circ$ did not successfully keep the relative argument at $90^\circ$ in the desired intensities above. Because the error of the relative argument and the decrease in the modulation amplitude hinder the improvement of the contrast, careful selection of the modulation area and its amplitude is required. The modulation EF should be arranged while considering observational conditions such as raw contrast, desired area size, target contrast improvement, and total integration time. The MWF for the arranged modulation EF will be added to the AO-corrected wave front in real observations. 
}

In the simulation, a perfect coherence of the speckle EF was assumed; however, the incoherent effect of bandwidth, even at 2\%, on the measurements should be studied in a future paper. The stability of the modulation amplitude, incoherent scattered light, and speckle (wave-front) variation characteristics should be studied in real optical conditions. On the other hand, compositions of multiple-band data might help increase the achievable contrast. Combining CDI-SAN with other processing methods might be also interesting. 
{A simulation using a phase screen model is in progress, which would be of advantage for the correlation problem. 
}
 
\section{Conclusion}
The CDI-SAN method was developed to detect faint exoplanets lying beneath varying residual speckles. CDI-SAN can be applied to both ground-based and space telescopes. It utilizes image acquisitions faster than stellar speckle variations synchronized repeatedly with five shapes of a deformable mirror. By using only the integrated values of each of the five kinds of images and several square differences for a long interval of observations, the light of the exoplanet can be separated from the stellar speckles. The achievable contrast would reach to almost the photon-noise limit of the residual speckle intensities under appropriate conditions, which would be helpful for exoplanet research missions today and in the future. 
\\

This work was partly supported by JSPS KAKENHI grant Nos. 19H01932 and 20H05893. The author would like to thank Dr. N.~Murakami for helpful discussions and Profs. S.~Tsuneta, Y.~Hayano, T.~Usuda, and M.~Tamura for constant encouragements. 


\begin{figure*}
\gridline{  \fig{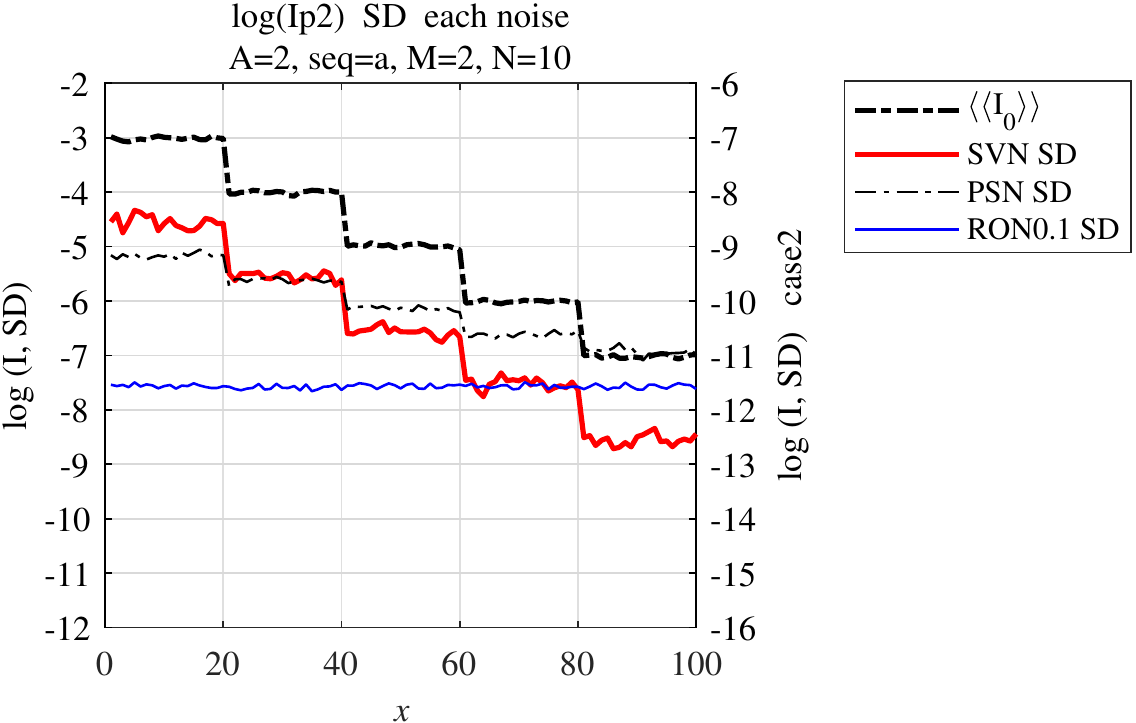}  {0.43\textwidth}{(a)}  		\fig{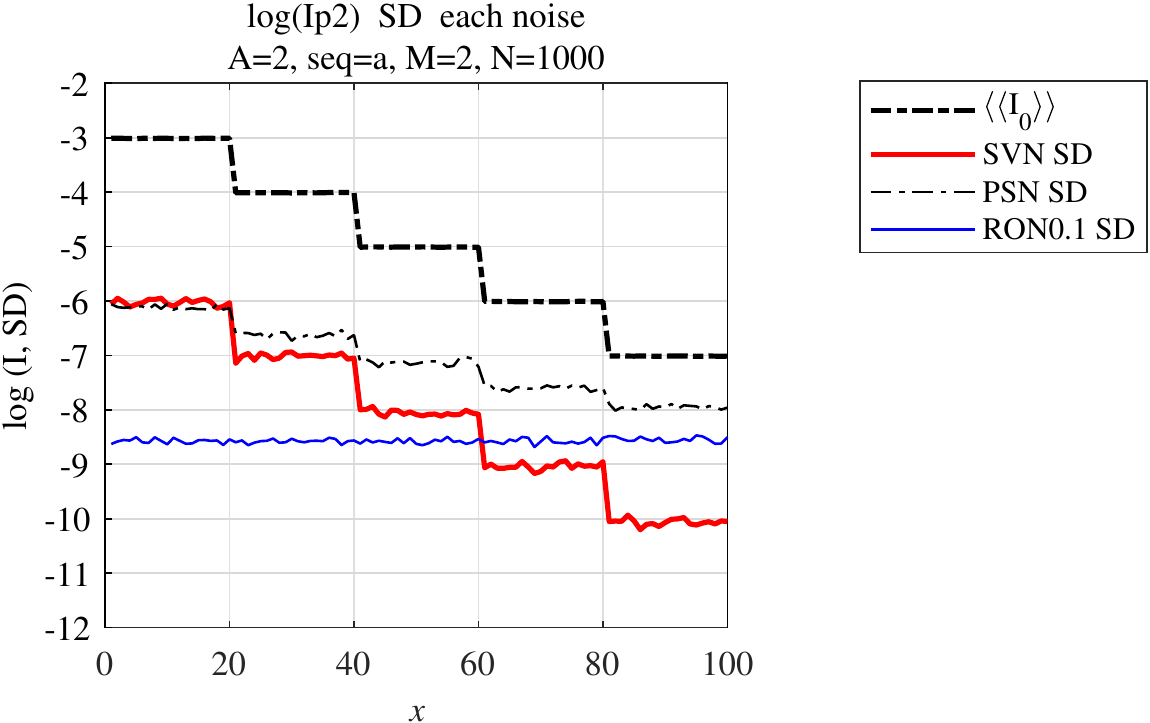}  {0.43\textwidth}{(c)}  }
\gridline{  \fig{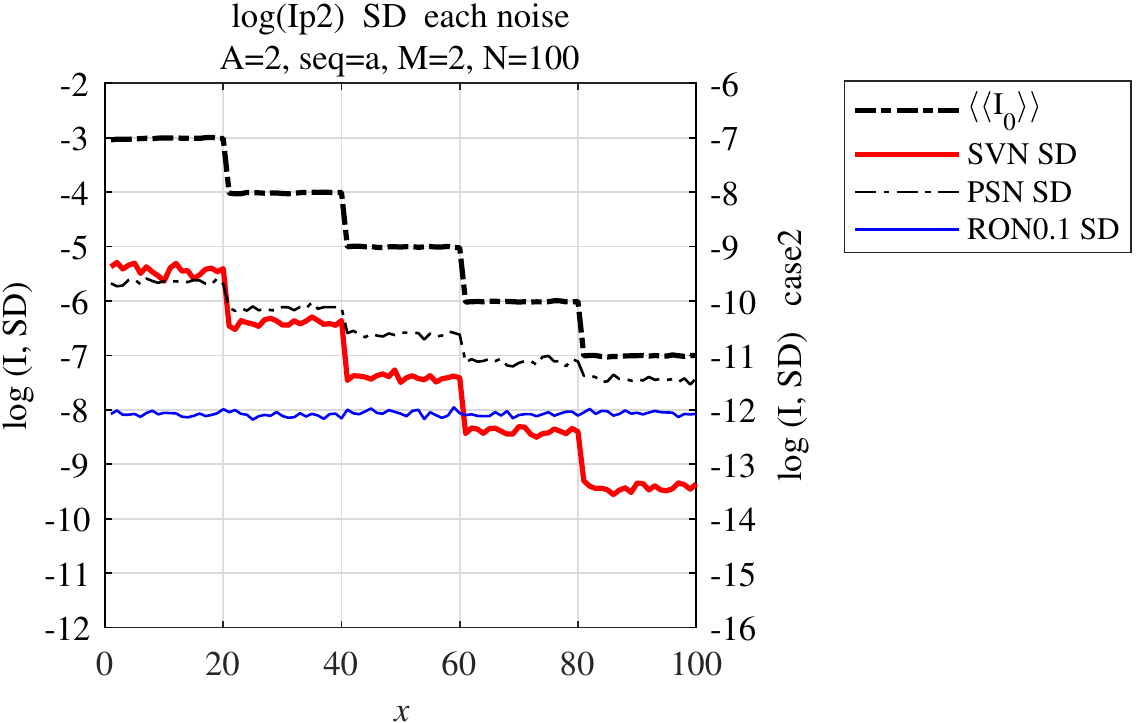}  {0.43\textwidth}{(b)}  		\fig{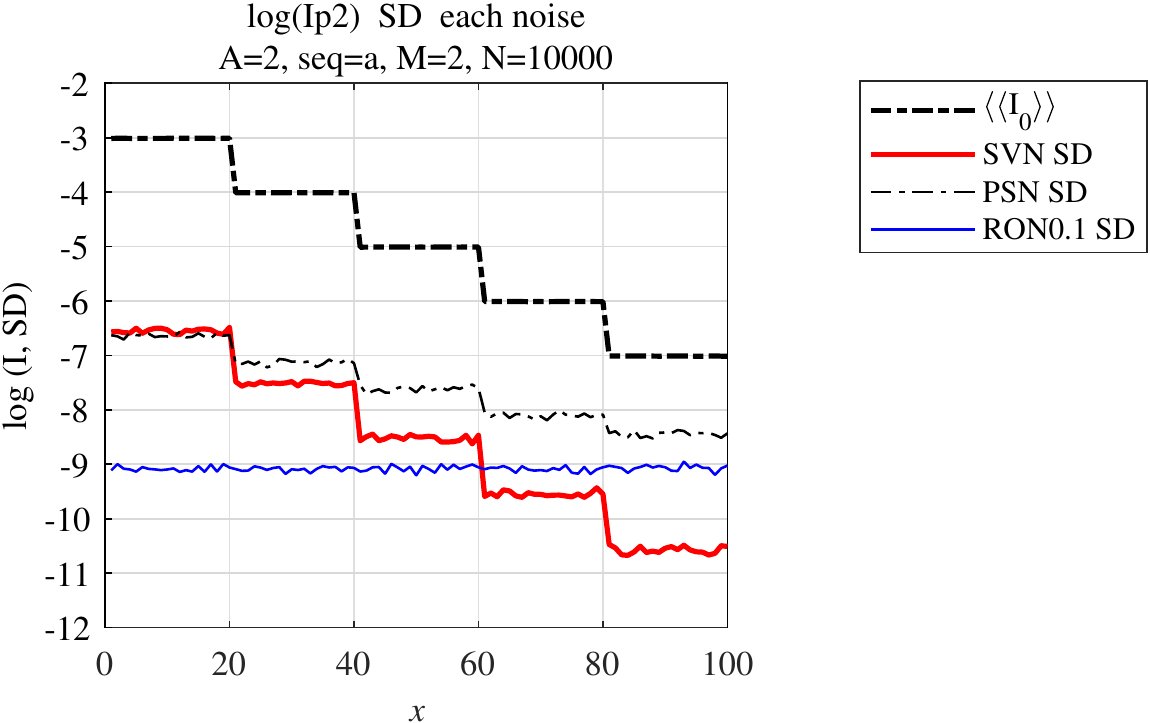}  {0.43\textwidth}{(d)}  }
\caption{Effects of the three individual noises in $I_{p2}$ shown by the SD$_x$ under $M$=2 as in Figure~\ref{Ip2CI} (c). (a) $N$=10, (b) $N$=100, (c) $N$=1000, and (d) $N$=10,000. Thick red line: SVN, dotted-dashed black line: PSN, solid blue line: RON(=0.1), thick dotted-dashed black line: $\langle\langle I_0 \rangle\rangle$ original contrast as a reference. The ordinate values are shown in log scale. The right ordinates in (a) and (b) show the contrast in the case~2. 
}
\label{Ip2CIN10}
\end{figure*}

\begin{figure*}
\gridline{  \fig{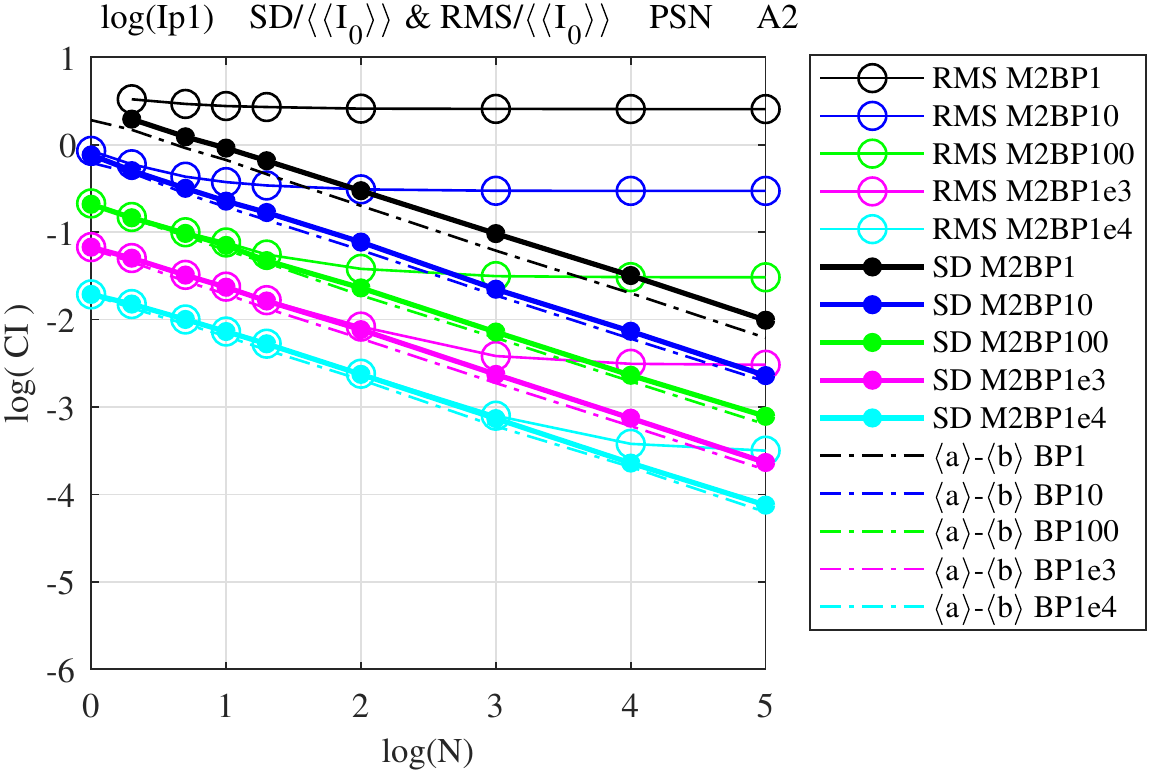}{0.43\textwidth}{(a)}        \fig{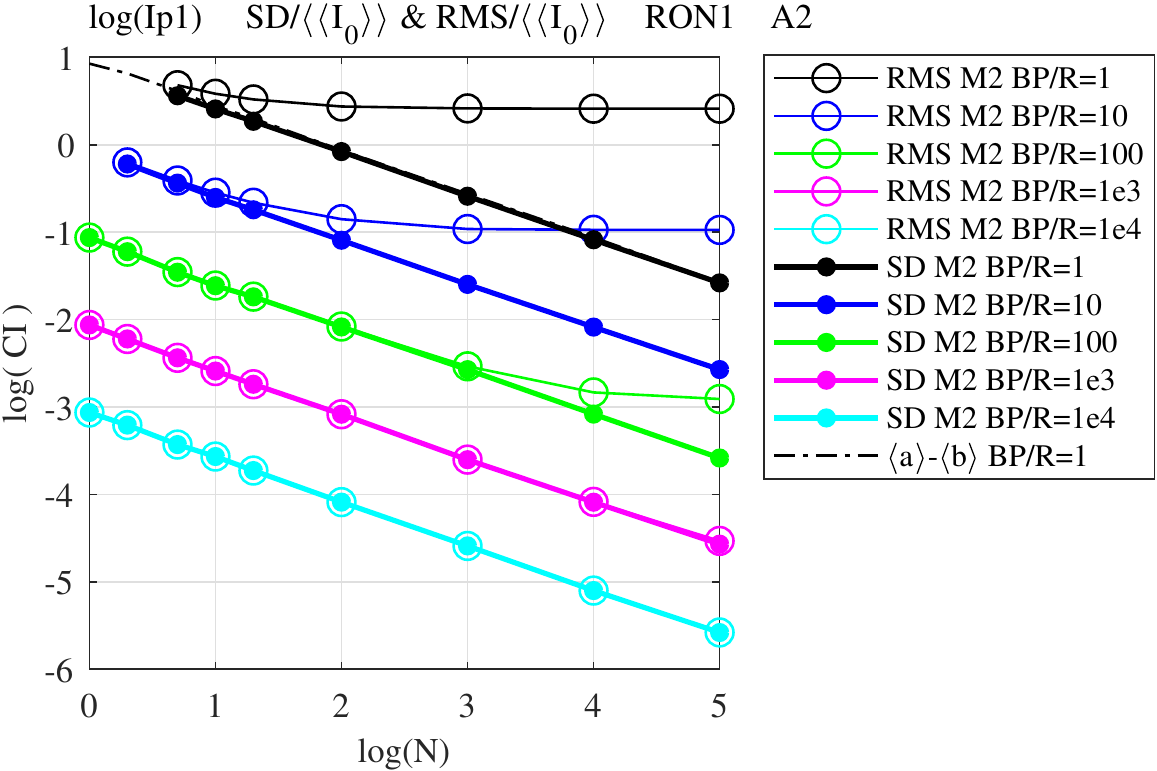}{0.43\textwidth}{(b)}  }
\caption{Contrast improvement of $I_{p1}$ for the two noises as a function of $N$. (a) PSN with $M$=2. Thick lines with filled circles: SD/$\langle\langle I_0 \rangle\rangle$, which was almost the same as $I_{p2}$, solid lines with open circles: RMS/$\langle\langle I_0 \rangle\rangle$, and (thin dot-dashed lines) PSN-limit lines calculated frob $(\langle a\rangle-\langle b\rangle)/\langle\langle I_0 \rangle\rangle$. (b) RON(=1) with $M=2$. Thick lines with filled circles: SD/$\langle\langle I_0 \rangle\rangle$, which was almost the same as $I_{p2}$, solid lines with open circles: RMS/$\langle\langle I_0 \rangle\rangle$, and dot-dashed line: the RON-limit line with $BP/R$=1 by $(\langle a\rangle-\langle b\rangle)/\langle\langle I_0 \rangle\rangle$. The lines in both panels were for $BP$ = 1, 10, 100, 1000, and 10,000 from top to bottom. 
}
\label{Ip1CI}
\end{figure*}

\appendix
\section{The SD$_x$ of $I_{p2}$ for Various \it{N}}
\label{VariousN}
Figure~\ref{Ip2CIN10} showed the SD$_{x}$ of $I_{p2}$, as in the Figure~\ref{Ip2CI} (a), with one of the SVN, the PSN, or the RON individually for $N$=10, 100, 1000, and 10,000 in (a), (b), (c), and (d), respectively, where $M$=2 and $R=0.1$ were used. The CIs are understood to be the SD$_{x}$ levels relative to ${\langle\langle I_0 \rangle\rangle}$ at the corresponding regions. The left ordinate indicated the results for the case~1. Figure~\ref{Ip2CIN10} (a) and (b) also show the contrast in the case~2 in the right ordinates. The CIs in this case~2 are the same as those in the case~1 because they depended on $P$ instead of $C$, where only $N$ should be $\la 100$ if $t_0$=3600s and a limited total integration time of 100 hr.

\section{characteristics of $I_{p1}$}
\label{Ip1}
 The characteristics of $I_{p1}$, Equation~(\ref{eqIp1}), are presented here. The solution $I_{p1}$ show the bias intensities of the PSN and the RON at low-flux regions, which are seen in saturations of the RMS/${\langle\langle I_0 \rangle\rangle}$ in contrast to $I_{p2}$ of Equation~(\ref{eqIp2}). They remain because the second and the third terms of the equation of $I_{p1}$ are larger than those without the noises, and then the uncorrected biases of $I_{p1}$ by the two noises are negative. 
Figure~\ref{Ip1CI}~(a) shows the rms/${\langle\langle I_0 \rangle\rangle}$ of $I_{p1}$ for the PSN, which saturated at a much larger level than the SD/${\langle\langle I_0 \rangle\rangle}$ levels at all regions. 
Figure~\ref{Ip1CI}~(b) showed the rms/${\langle\langle I_0 \rangle\rangle}$ saturations of $I_{p1}$ for the RON, which were found specifically at the low-photon-flux regions. 

However, the SD/${\langle\langle I_0 \rangle\rangle}$ for the PSN shown in Figure~\ref{Ip1CI}~(a) was almost the same as $I_{p2}$ in Figure~\ref{Ip2CI}~(b), about 1.2 times ${(\langle a \rangle-\langle b \rangle)/\langle\langle I_0 \rangle\rangle}$ at $BP\geq10$ between $10\le N\le100,000$. At a very low photon flux of $BP$=1, the increase is still only about 1.6 times at $M$=2 and gradually rises with $M$, which is slightly better than $I_{p2}$. 
The SD/${\langle\langle I_0 \rangle\rangle}$ for the RON ($R$=1) in Figure~\ref{Ip1CI}~(b) was almost the same as $I_{p2}$ in Figure~\ref{Ip2CI}~(c) and close to the RON limit of ${(\langle a \rangle-\langle b \rangle)/\langle\langle I_0 \rangle\rangle}$ at $50\le N\le100,000$. The SD of $I_{p1}$ against the RON can be improved by $1/\sqrt{2}$ by not dividing the five intensities into two half-{exposures}, if available. 

The biases should be subtracted in some way to know the intensity of the planets. 
If the bias intensities of the PSN and the RON can be removed by a smooth surface at the focal plane, as in the SVN case shown in Figure~\ref{Ip2image} (c) and (d) in a real optical condition, we do not have to use $I_{p2}$ to remove the biases but can use $I_{p1}$, which has slightly better SDs in some conditions and has the advantage of slow data acquisition without splitting into two half-{exposures}. If surface subtraction is not available, it would be good to use $I_{p2}$ to escape from the bias problem. 
It was found that the biases in $I_{p1}$ from the RON could be reduced with a large $A$, and the large $A\le10$ did not affect the SDs of the three noises by more than 20\% in the simulation. Large $A$ introducing bright modulation intensities, however, might have a risk of additional measurement errors in practice.  

On the other hand, on the SVN,  the RMS/${\langle\langle I_0 \rangle\rangle}$ and the SD/${\langle\langle I_0 \rangle\rangle}$ of $I_{p1}$ were almost the same as those of $I_{p2}$ shown in Figure~\ref{Ip2CI} (b). The rms of  $I_{p1}$ was slightly small, about 0.9 times, $I_{p2}$ at $1000\le N\le100,000$.  For either $I_{p1}$ or $I_{p2}$ bias intensities exist, as shown in Figure~\ref{Ip2image} (b), which should be removed by the smooth surface as shown in Figures~\ref{Ip2image} (c) and (d) or by a large $M$. The contrast achievable by $I_{p1}$ and $I_{p2}$ should be studied in real optical conditions. 

\begin{figure}
\gridline{	\fig{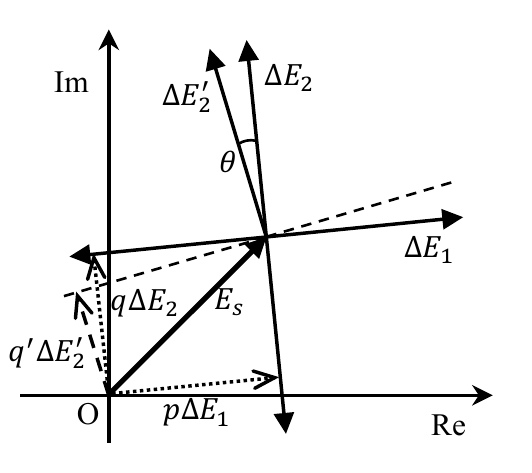}  {0.28\textwidth}{}  } 
\caption{Electric field at a pixel with the relative argument error  $\theta$.}
\label{EFArgErr}
\end{figure}

\section{Free-argument solution}
\label{FreeArgument}
{
The solutions, equations~(\ref{eqIsA}) and (\ref{eqIp1}), were derived by assuming the orthogonality of the $\Delta E_1$ and $\Delta E_2$. The error of the relative argument between them, $\theta$, seen in Figure \ref{EFArgErr}, would produce an increase in the SD. When $\theta$ is known, given by a model, it would be possible to use a solution without the orthogonality assumption. By using  $\Delta E_2^{'}$ with $\theta$ instead of $\Delta E_2$, the solution can be written as }

\begin{equation}
\label{eqI12A2}
\left\{
\begin{array}{l} 
\langle I_2^{'+} \rangle =\langle \left|E_s+\Delta E'_2\right|^2 \rangle +I_p \\
\langle I_2^{'-} \rangle =\langle \left|E_s-\Delta E'_2\right|^2 \rangle +I_p , \\ 
\left|\Delta E'_2\right|^2\ ={\left(\langle I_2^{'+}\rangle +\langle I_2^{'-}\rangle  -2\langle I_0\rangle \right)/2} \\
\langle (\Delta E'_2\bullet E_s)^2\rangle\ =\langle\left(I_2^{'+}\ -I_2^{'-}\right)^2\rangle/16  \\
q' =\frac{\Delta E'_2\bullet E_s}{\left|\Delta E'_2\right|^2\ }=\frac{\left(I_2^{'+}-I_2^{'-}\right)}{2\left(I_2^{'+}+I_2^{'-}-2I_0\right)},\\
q =\frac{q'}{\cos{\theta}}+\frac{p\left|\Delta E_1\right|\sin{\theta}}{\left|\Delta E'_2\right|\cos{\theta}} \\
\end{array}
\right. 
\end{equation}

\begin{eqnarray}
\label{eqIsA2}
\langle I_{s} \rangle & = & \frac{\langle ( \Delta E_1 \bullet E_s )^2 \rangle}{\left| \Delta E_1 \right|^2 \cos^2{\theta} } 
+\frac{2\langle ( \Delta E_1\bullet E_s ) ( \Delta E'_2\bullet E_s ) \rangle \sin{\theta}}{\left|\Delta E_1\right|\left|\Delta E'_2\right|\cos^2{\theta} } \nonumber 
+\frac{\langle(\Delta E'_2\bullet E_s )^2\rangle}{\left|\Delta E'_2\right|^2\cos^2{\theta} } \nonumber \\
 & = & 
   \frac{\langle\left(I_1^+-I_1^-\right)^2\rangle }{8\left( \langle I_1^+ \rangle + \langle I_1^- \rangle -2 \langle I_0 \rangle \right) \cos^2{\theta}} 
+ \frac{\langle\left(I_1^+-I_1^-\right)\left(I_2^{'+}-I_2^{'-}\right)\rangle \sin{\theta}}
{4\sqrt{ \left( \langle I_1^+ \rangle + \langle I_1^{'-} \rangle -2 \langle I_0 \rangle \right) \left( \langle I_2^{'+} \rangle + \langle I_2^{'-} \rangle -2 \langle I_0 \rangle \right) } \cos^2{\theta}} \nonumber \\
&+& \frac{\langle\left(I_2^{'+}-I_2^{'-}\right)^2 \rangle }
{8\left( \langle I_2^{'+} \rangle + \langle I_2^{'-} \rangle -2 \langle I_0 \rangle \right) \cos^2{\theta}}  \ . 
\end{eqnarray}


\end{document}